# Characterization of the Si(Li) detector for Monte Carlo calculations of beta spectra

**Pavel Novotny[a,b], Pavel Dryak[a], Jaroslav Solc[a,*], Petr Kovar[a], and Zdenek Vykydal[a]**

[a] *Czech Metrology Institute, Okruzni 31, 638 00 Brno, Czech Republic*
[b] *Faculty of Nuclear Sciences and Physical Engineering, Czech Technical University in Prague, Brehova 78/7, 115 19 Prague 1, Czech Republic*
[*] *corresponding author; e-mail*: jsolc@cmi.cz

ABSTRACT: A precise model of a Si(Li) detector ORTEC® model SLP-06165P-OPT-0.5 was created for beta spectra calculations using the Monte Carlo (MC) code MCNPX™. Detector parameters were determined from X-ray radiograms obtained with a film and a Timepix detector. The MC model of the detector was validated by comparison of calculated and experimental full-energy peak efficiencies in the energy range from 5 to 136 keV using a range of point-like photon emitting radionuclide standards. A comparison of measured and calculated beta spectrum of a radionuclide Pm-147 is presented.

## Contents

## 1. Introduction

Motivation for the presented work is to adapt a routine method that enables identification and quantification of pure beta impurities in pure beta radionuclides resulting in the reduction of minimum detectable activities, e.g., mixtures of Sr-89 and Sr-90, or Y-90 and Sr-90. This is very important especially for absolute activity measurement by liquid scintillation counting, or using radiopharmaceutical Y-90 for PET diagnosis, where Sr-90 impurity would be very dangerous for human body. Namely in nuclear medicine, determination of radiochemical purity is required by TRS 454 [1].

There are various methods for measurement of beta spectra but the majority of them, e.g. magnetic spectrometry, cannot be used for routine laboratory measurements, as they are suited for research purposes and the measurement is expensive and time consuming. For routine measurements we continued the work published in [2] and selected a Lithium-Drifted Silicon Low-Energy Photon Spectrometer (Si(Li) spectrometer). With a sufficient accuracy it allows determination of the end point that is, together with the spectrum shape, the main parameter of a continuous beta spectrum.

Comparison of beta spectrum obtained experimentally with the one determined by a Monte Carlo (MC) simulation aims to find the ratio of activity of different pure beta radionuclides in mixtures. To achieve this, a MC model of a Si(Li) detector used in a spectrometric laboratory of the Czech Metrology Institute (CMI) was created for beta spectra calculation. The correctness of the detector model was verified using point-like standard sources with X-ray and $\gamma$-ray emitting radionuclides produced by CMI. The verified MC model was then used for preliminary comparison of simulated and measured beta spectrum.

Characterization of a Si(Li) spectrometer has been reported by several authors, e.g. [3] and [4], exploiting different techniques. In the presented paper, description of the detector was performed by using X-ray radiography with a film and with a Timepix detector. Compared to the work [2], this work describes a new piece of the detector after the replacement of the former one. Structure of the new detector was analysed by additional methods and the developed Monte Carlo model is more precise resulting in significantly better agreement between measured and simulated photon detection efficiencies and beta spectra.

## 2. Materials and methods

### 2.1 Si(Li) Spectrometer

The CMI's spectrometric laboratory uses an SLP Series Lithium-Drifted Silicon Low-Energy Photon Spectrometer model SLP-06165P-OPT-0.5 (Ortec®;[5]) for both X-ray spectrometry and beta spectrometry. Nominal detector parameters are as follows: active diameter 6 mm, sensitive depth 5 mm, absorbing layers: Be 0.0127 mm, Au 20 nm, and Si 0.1 mm. Resolution at 5.9 keV is 160 eV (full-width at half maximum). The detector is installed inside a 5 cm thick lead shielding. For measurement of X-ray and γ-ray sources, a stand made of acrylic glass is inserted into the shielding fixing the measurement position of the sources. For measurement of beta sources, the acrylic glass stand is replaced with a special collimator and source holder made of aluminium alloy. Its presence significantly reduces scattered radiation and allows a better description of the beta particle beam. Diameter of the collimator aperture is 4 mm.

Typical beta sources measured with this detector are produced by dropping a radionuclide solution onto a 0.1 mm thick polyethylene terephthalate (PET) foil. Source activity ranges from 50 to 100 kBq. Distance between the source and the detector end cap is 35 mm.

### 2.2 Confirmation of the spectrometer parameters

Description of the crystal and the construction elements provided by the detector manufacturer was confirmed using X-ray radiography with a defectoscopic film sheet (Structurix D5 Pb Vacupac, AGFA) and a Timepix detector equipped by 300 μm thick Si sensor. Timepix is a single photon counting semiconductor pixel detector with 256×256 square pixels 55 μm in size [6]. The sensitive area is about 2 $cm^2$, so the Si(Li) detector had to be imaged in several steps. The radiography also allowed to determine other parameters not provided by the manufacturer, and to obtain an accurate picture of the detection system inner construction, especially the shape of the crystal and its position within the cap, and the cryostat structure. The film was exposed to 60 kV RQR4 X-ray radiation quality for 280 s delivering the total air kerma of 4 mGy at a distance of 1.5 m from the X-ray source. The Timepix detector was exposed to unfiltered 150 kV X-ray beam. Each imaging sequence took 100 s. The response of each individual pixel was calibrated by direct thickness calibration method using Al filters up to 52 mm thick [7].

### 2.3 Monte Carlo model

The MC model consisted of the Si(Li) detector and its cryostat, collimator, shielding, and an appropriate source on a stand (Figure 1). MC simulations were performed using the general-purpose MC code MCNPX™ in version 2.7.E [8]. Continuous-energy photoatomic data library MCPLIB84 [9] and electron condensed-history library el03 with 1 keV and 10 keV energy cut-off were used for photon and electron transport, respectively. The el03 tables are based on the Integrated TIGER Series 3.0 and their description is provided in [10].

For beta spectra calculation only, to sufficiently sample electron energy and angular straggling in materials between the source and the detector in condensed-history electron transport, values of two parameters influencing the straggling were modified – efac in PHYS:E card and ESTEP on material card. The efac controls stopping power energy spacing and it was changed from the default value of 0.917 to 0.96 [8]. The parameter ESTEP defines the number of electron sub-steps per energy step [8] and it was increased to 30, 200, and 100 in sensitive silicone, insensitive silicone, and gold, respectively. Also, ESTEP value was increased in the air and in the material of the source. Their value depended on the end point of a calculated beta spectrum.

In addition, the efficiency of beta spectrum calculation was increased by implementation of two variance reduction methods: cell-by-cell electron energy cut-off and source bias. Cell-by-cell energy cut-off for electrons was used for aluminium alloy outer collimator, except for its small part close to an aperture, by setting an ELPT card [8] to 300 keV. This method suppressed tracking of low energy electrons in parts of the geometry from which they cannot contribute to the detector spectrum. The exponential source bias with the parameter a=1 [8] allowed to emit the electrons from the source preferentially towards the detector increasing the probability to contribute to the detector spectrum.

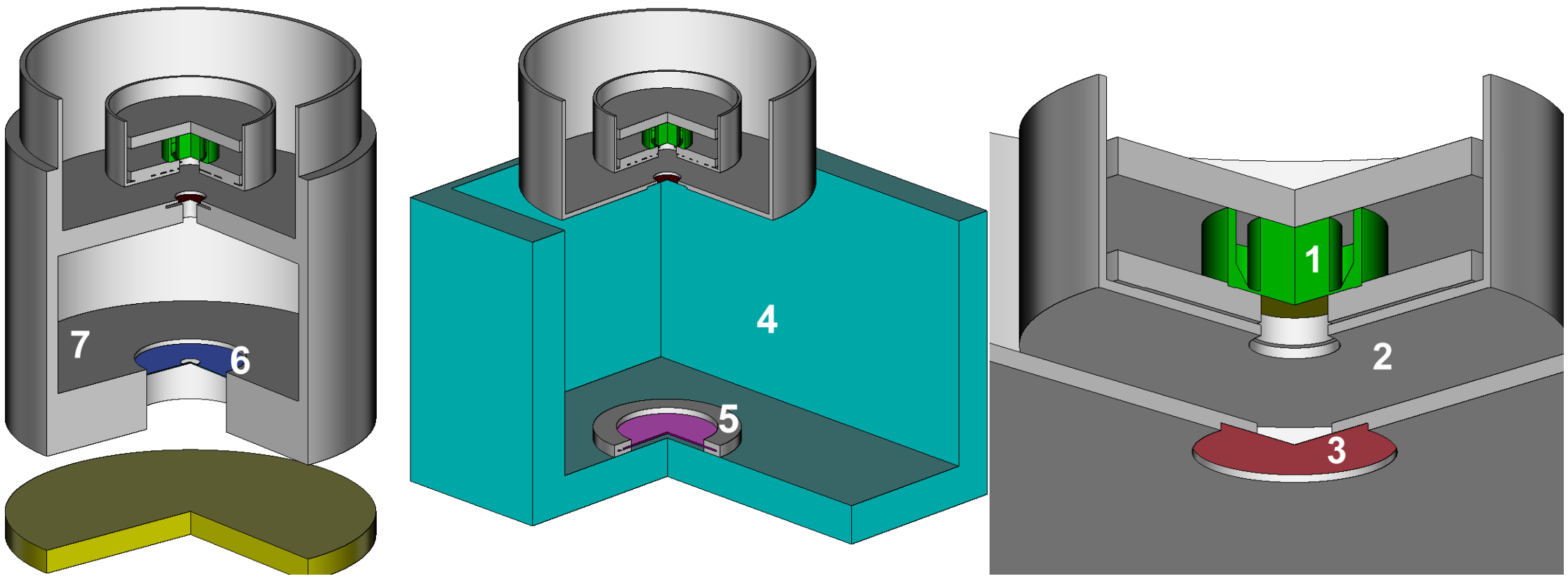

**Figure 1**. Visualizations of the MC model of the Si(Li) detector. Left - set-up for beta spectra measurement; centre – set-up for MC model validation; right – detail of the detector. Colours distinguish different materials. Main parts of the set-up are numbered as follows: 1 – Si crystal, 2 – crystal cover and inner collimator, 3 – Be entrance window in the cryostat cover, 4 – acrylic glass stand, 5 – photon source, 6 – beta source, 7 – aluminium alloy holder of beta sources and outer collimator.

### 2.4 MC model validation

The MC model of the Si(Li) detector was validated by comparison of experimental and calculated full-energy peak (FEP) detection efficiencies. Experimental FEP efficiencies were determined in the energy range from 5 to 136 keV using point-like sources with a range of X-ray and γ-ray emitting radionuclides (see Table 1) prepared in CMI and traceable to the Czech national standard for activity of radionuclides. The calculated FEP efficiencies were obtained using the detector pulse-height tally of type F8 [8] for monoenergetic source photons of the energy matching the measured sources. The particles were emitted from the source into a cone oriented towards the detector crystal to increase computational efficiency. The outer aluminium collimator was removed for these measurements. The sources were placed on an acrylic glass stand and positioned on the detector axis at a distance of 63.5 mm from the end cap. The part of the MC model outside the cryostat was modified appropriately to match the geometry of the measurement. The MC simulation was stopped when the statistical standard uncertainty of the FEP efficiency reached 0.2%.

### 2.5 Beta spectrum comparison

The preliminary comparison of a measured and calculated beta spectrum was performed for the pure beta emitter Pm-147. The beta spectrum of Pm-147 has the end point of 224.1 keV and the yield of 1.00 electrons per decay [11]. The source was prepared at CMI by dropping an aqueous solution of Pm-147 onto a PET foil creating a very thin source with a radius of 1.5 mm. The source activity was 18.77 kBq (±1.5% [12]) at the reference date of 11 October 2017. The beta spectrum acquisition time was 18.34 h.

In the simulation, the Pm-147 emission spectrum was taken from the beta spectra compilation available at [13]. The source was modelled as a cylinder with 50 µm height, 1.5 mm radius, the density of 0.1 g/cm$^3$, and the elemental composition $PrCl_3$ + $NdCl_3$ [12]. The ESTEP value in air and in the source was set to 10 and 15, respectively, resulting in average 52 energy sub-steps in air between the source and the detector window.

## 3. Results

### 3.1 Parameters of Si(Li) spectrometer

Figures 2 and 3 show a radiogram of the Si(Li) detector obtained with the film and the Timepix detector, respectively. The radiogram from the Timepix is composed of 15 individual exposures. Both radiograms clearly show shapes of individual parts of the detector that is necessary for the development of a precise MC model. A schematic drawing of the Si(Li) detector with parameters obtained from the radiograms and then used in the MC model is presented in Figure 4. The following thicknesses of insensitive volumes of the Si crystal were defined in the MC model: 500 µm at the rear, 175 µm on the side, and 0.1 µm at the front. In addition, there is an additional attenuation layer at the Si front made of 0.02 µm of gold foil.

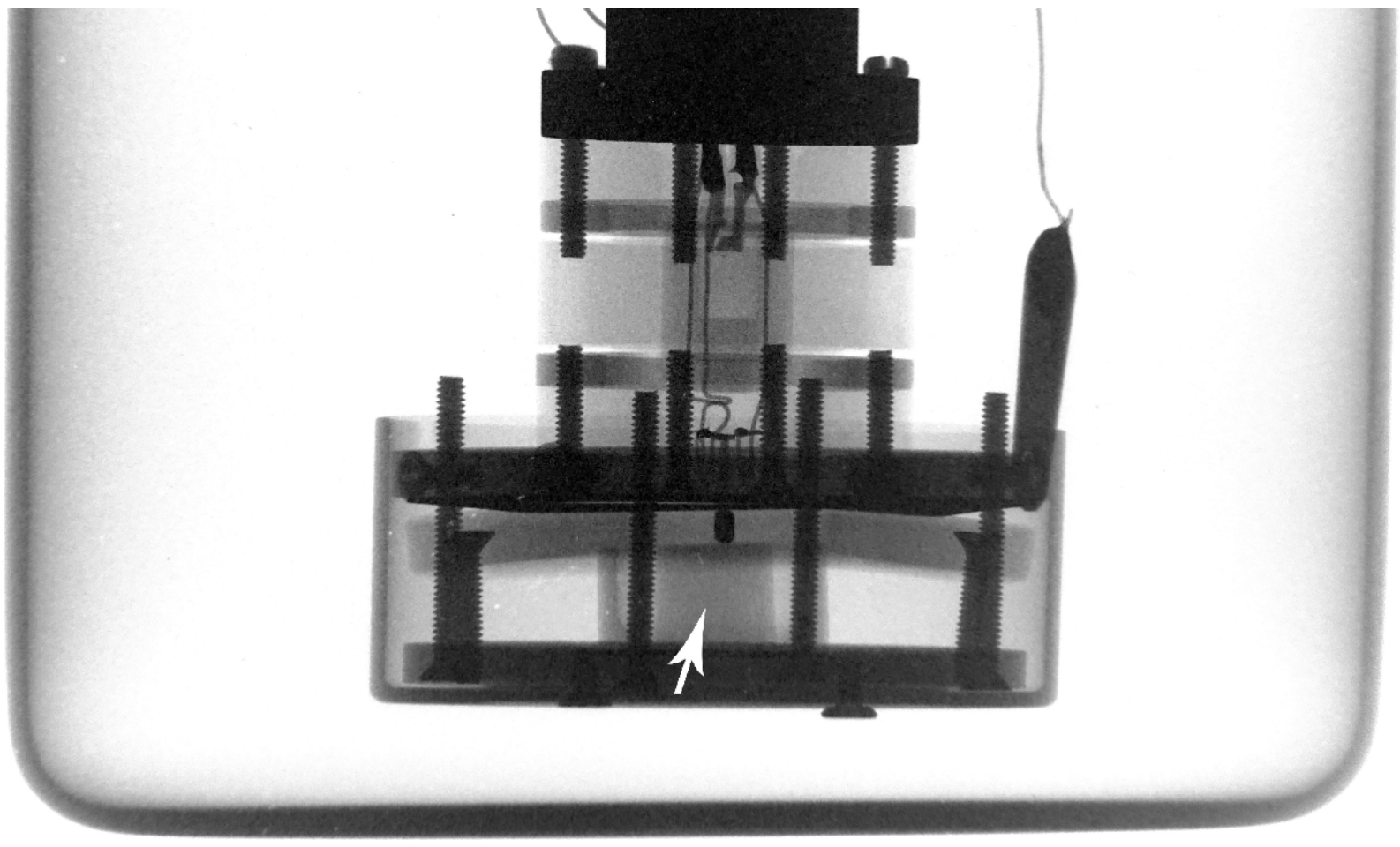

**Figure 2.** Si(Li) detector radiogram obtained with a film. Image is in inverse colours. The arrow points to the Si crystal.

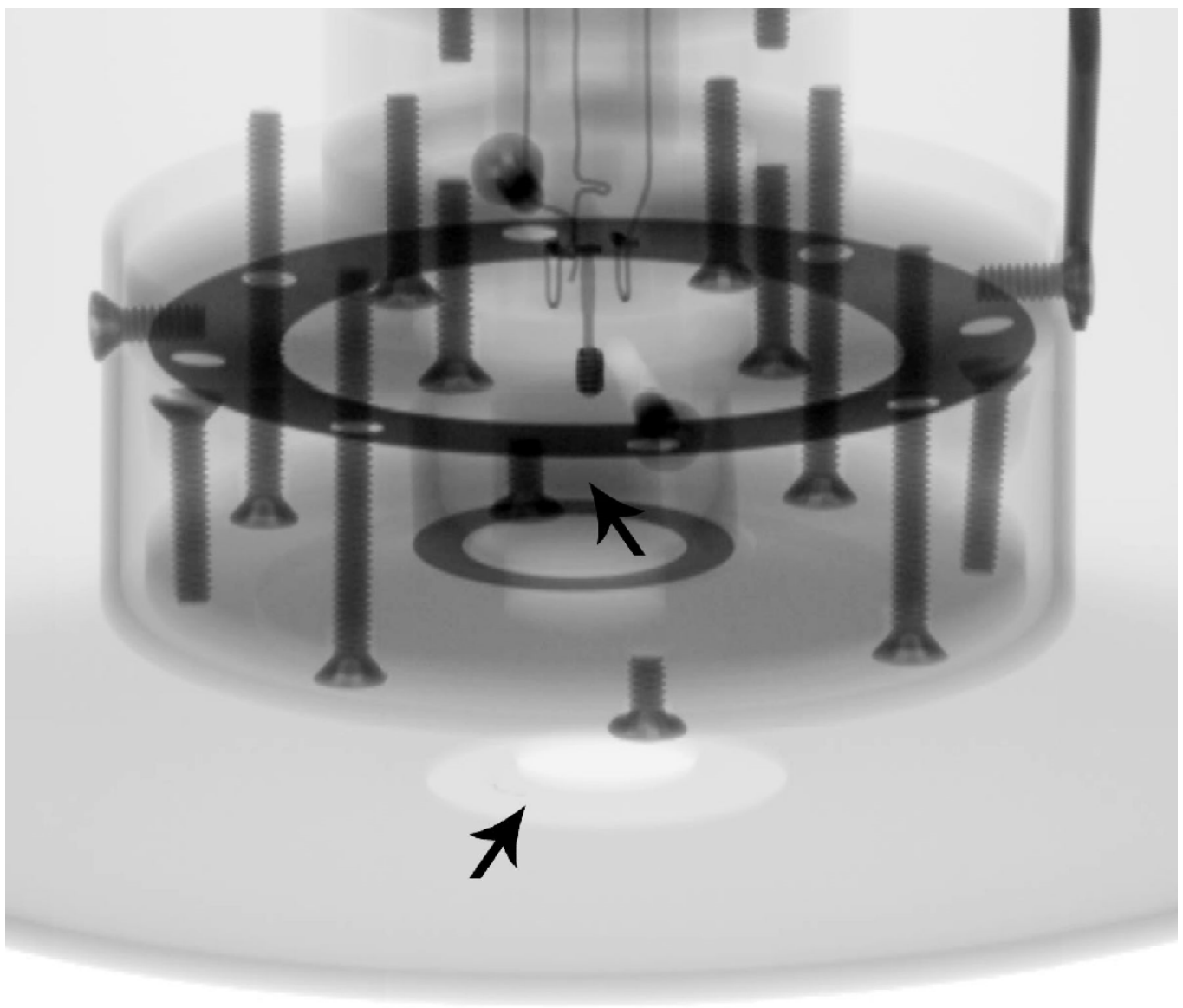

**Figure 3.** Si(Li) detector radiogram obtained with a Timepix detector. Image is in inverse colours. The arrows point to the entrance window in the end cap (the bottom one) and to the Si crystal.

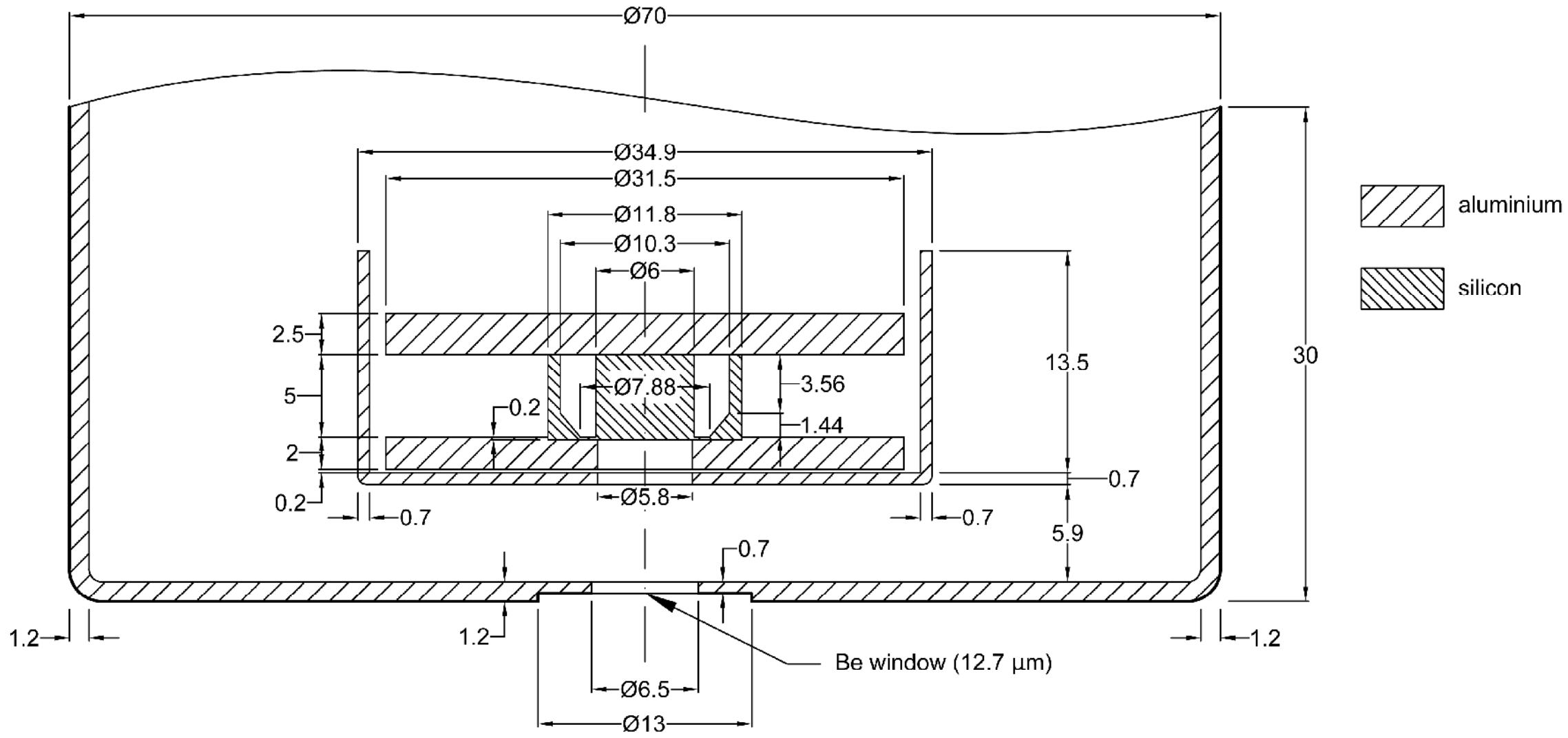

**Figure 4.** A schematic drawing of the Si(Li) detector and dimension obtained from radiograms and used in the MC model. Dimensions are given in millimetres.

### 3.2 MC model validation

The net peak areas were calculated by the total peak area method with a step function for continuum subtraction [14]. True coincidence summing corrections were negligible. The combined standard uncertainty of the experiment consists of the uncertainty of the net peak area, the source activity, and the photon yield. All these components are presented in Table 1. The uncertainty of the photon yield was taken from [11]. The standard uncertainties for dead-time measurement and for random summations were negligible. The standard uncertainty of the relative difference between experiment and simulation was obtained according to [15], section 5.1.2, assuming no correlation between uncertainty components.

Results of the comparison of the calculated and the experimental values of the FEP efficiencies are summarized in Table 1 and visualized in Figure 5. For all measured photon energies, relative differences vary within ±5% which the authors consider to be an acceptable result with respect to uncertainties of the experimental efficiency values. Therefore, the MC model of the Si(Li) detector is validated and can be used for beta spectra simulations.

**Table 1:** Comparison of experimental ($\eta_E$) and calculated ($\eta_C$) values of full-energy peak efficiency. *E* is photon energy, *u(A)*, *u(Y)*, and *u(S)* are experimental relative standard uncertainty of source activity, photon yield, and peak area, respectively, $u(\eta_E)$ is experimental relative combined uncertainty, and *RD* is a relative difference obtained as $RD = \eta_C/\eta_E - 1$. Calculated relative standard uncertainty is always 0.2% and it consists of the statistical uncertainty of the calculation only.

| Nuclide | Line | E (keV) | $\eta_E$ | u(A) | u(Y) | u(S) | u($\eta_E$) | $\eta_C$ | RD |
|---|---|---|---|---|---|---|---|---|---|
| Mn-54 | Cr Ka | 5.41 | 2.122E-04 | 1.0% | 5.3% | 0.3% | 5.4% | 2.192E-04 | (3.3 ± 5.6)% |
| Co-57 | Fe Kα | 6.40 | 2.732E-04 | 1.2% | 1.5% | 0.1% | 1.9% | 2.710E-04 | (-0.8 ± 1.9)% |
| Co-57 | Fe K'β1 | 7.06 | 2.935E-04 | 1.2% | 1.9% | 0.3% | 2.2% | 2.939E-04 | (0.1 ± 2.3)% |
| Pb-210 | Bi Lα1 | 10.84 | 3.581E-04 | 1.5% | 5.8% | 0.2% | 6.0% | 3.479E-04 | (-2.8 ± 5.8)% |
| Am-241 | Np La1 | 13.95 | 3.593E-04 | 1.9% | 1.0% | 0.6% | 2.2% | 3.575E-04 | (-0.5 ± 2.2)% |
| Co-57 | γ | 14.41 | 3.671E-04 | 1.2% | 1.6% | 0.2% | 2.0% | 3.579E-04 | (-2.5 ± 2.0)% |
| Cd-109 | Ag Kα2 | 21.99 | 3.488E-04 | 1.5% | 3.7% | 0.6% | 4.1% | 3.452E-04 | (-1.0 ± 4.0)% |
| Cd-109 | Ag Kα1 | 22.16 | 3.593E-04 | 1.5% | 3.6% | 0.3% | 3.9% | 3.440E-04 | (-4.3 ± 3.7)% |
| Ba-133 | Cs Kα2 | 30.63 | 2.467E-04 | 1.0% | 2.6% | 0.1% | 2.8% | 2.478E-04 | (0.5 ± 2.8)% |
| Ba-133 | Cs Kα1 | 30.97 | 2.372E-04 | 1.0% | 1.1% | 0.1% | 1.5% | 2.432E-04 | (2.6 ± 1.5)% |
| Eu-152 | Sm Kα2 | 39.52 | 1.487E-04 | 1.0% | 2.4% | 0.4% | 2.6% | 1.471E-04 | (-1.1 ± 2.6)% |
| Eu-152 | Sm Kα1 | 40.12 | 1.362E-04 | 1.0% | 2.6% | 0.3% | 2.8% | 1.420E-04 | (4.2 ± 2.9)% |
| Am-241 | γ | 59.54 | 4.603E-05 | 1.9% | 1.1% | 0.4% | 2.2% | 4.823E-05 | (4.8 ± 2.4)% |
| Eu-152 | γ | 121.78 | 5.185E-06 | 1.0% | 0.2% | 3.1% | 3.3% | 5.233E-06 | (0.9 ± 3.3)% |
| Co-57 | γ | 122.06 | 5.468E-06 | 1.2% | 0.2% | 0.6% | 1.3% | 5.197E-06 | (-5.0 ± 1.3)% |
| Co-57 | γ | 136.47 | 3.777E-06 | 1.2% | 0.7% | 1.9% | 2.4% | 3.651E-06 | (-3.3 ± 2.3)% |

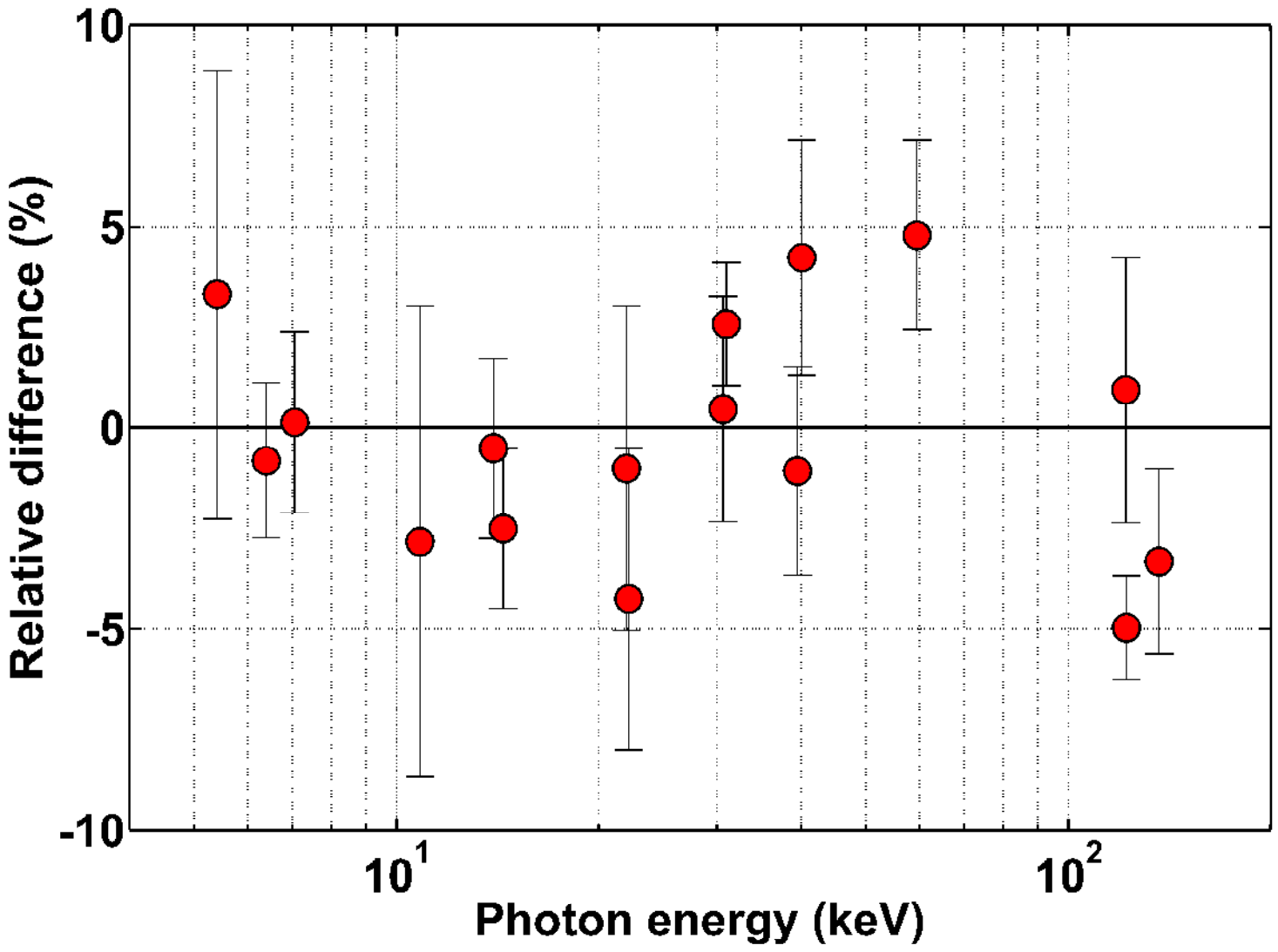

**Figure 5:** Relative difference (RD) between measured ($\eta_E$) and calculated ($\eta_C$) full-energy peak efficiencies for photons from 5 to 136 keV, determined as RD = $\eta_C/\eta_E$ -1.

### 3.3 Beta spectrum comparison

The preliminary comparison of the measured beta spectrum of Pm-147 with the one calculated with the validated MC model is presented on Figure 6. The spectra are compared absolutely in the number of counts per decay per energy interval of 2 keV width.

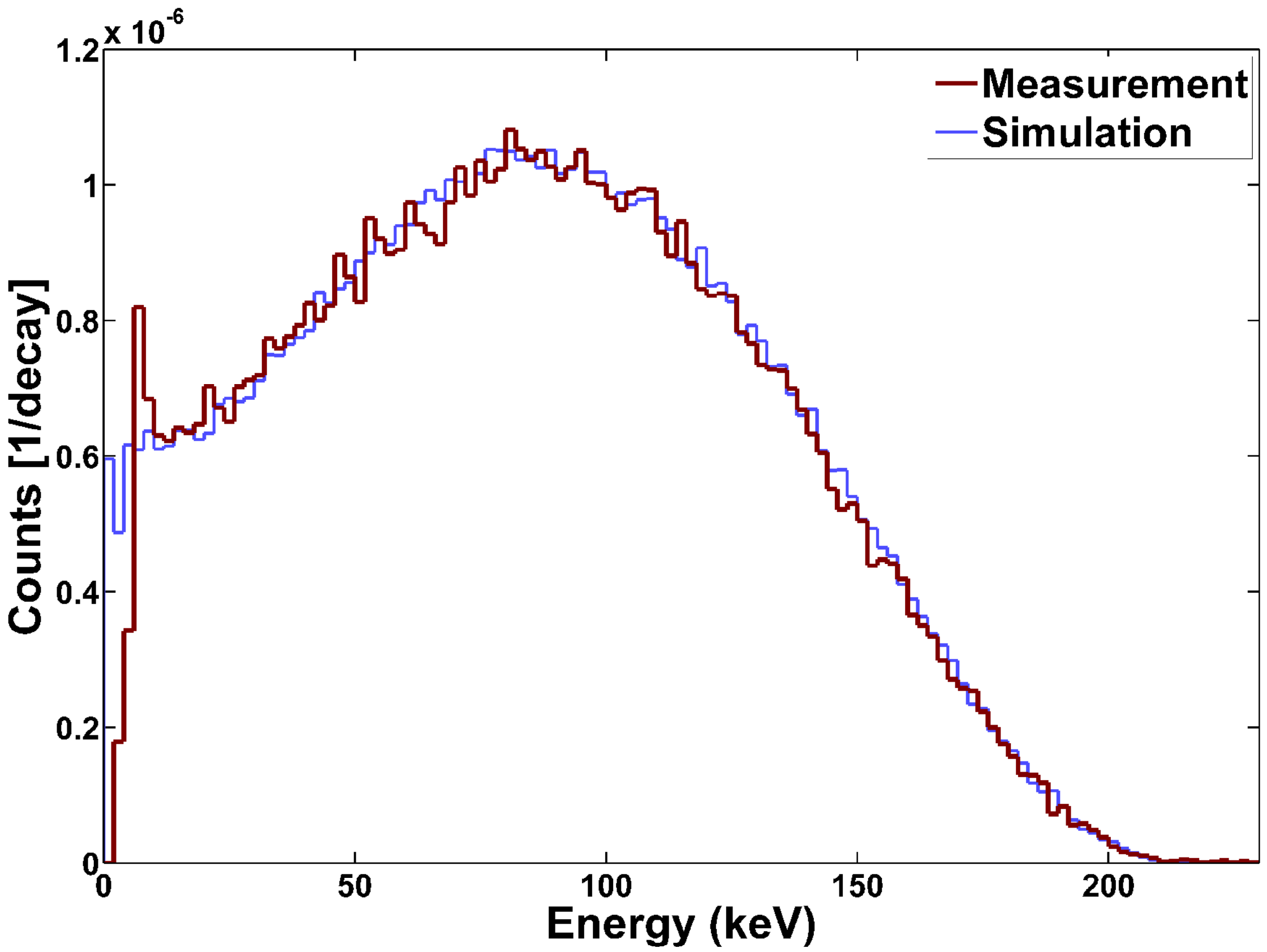

Figure 6: Comparison of measured (dark red) and calculated (light blue) beta spectrum of Pm-147 obtained with the Si(Li) detector.

The excellent agreement of the measured and simulated detector spectrum of Pm-147 confirms significant improvement of the beta spectra measurement and calculation originally published in [2] and prove the accuracy of the developed MC model of the Si(Li) detector and the whole geometry for the measurement of beta spectra.

## 4. Conclusions

A Si(Li) spectrometer was characterized by X-ray radiography and the obtained detector parameters were used for preparation of an MCNPX™ input file usable for Monte Carlo calculations of beta spectra. The MC model was validated using a set of point-like standard sources with X-ray and γ-ray emitting radionuclides. Measured and calculated full-energy peak efficiencies agreed within ±5%. The accuracy of the MC model was demonstrated on the comparison of the calculated beta spectrum of a radionuclide Pm-147 with the one measured by the Si(Li) detector.

The MC model will be used for calculation of beta spectra mixtures and minimum detectable activities determination, especially for radionuclides Sr-89, Sr-90, Y-90, P-32 and P-33. The results will permit to decrease the uncertainty of absolute activity measurement of pure beta radionuclides and radionuclide impurities determination in radiopharmaceuticals.

## Acknowledgments

This work was supported by the European Metrology Programme for Innovation and Research (EMPIR) joint research project 15SIB10 "Radionuclide beta spectra metrology" (MetroBeta; http://metrobeta-empir.eu/) which has received funding from the European Union. The EMPIR initiative is co-funded by the European Union's Horizon 2020 research and innovation programme and the EMPIR Participating States.